# Bound state in the continuum and polarization-insensitive electric mirror in low-contrast metasurface

Hao Song[1,2], Xuelian Zhang[1,2], Jian Wang[1,2]*, Yanming Sun[3] and Guo Ping Wang[3]*

[1]*College of Physics and Electronic Information Engineering, Neijiang Normal University, Neijiang, 641112, P. R. China*

[2]*Photonics Center, Neijiang Normal University, Neijiang, 641112, P. R. China*

[3]*College of Electronics and Information Engineering, Shenzhen University, Shenzhen, 518060, P. R. China*

*Email: gpwang@szu.edu.cn

**Abstract**

High-contrast refractive indexes are pivotal in dielectric metasurfaces for inducing various exotic phenomena, such as the bound state in the continuum (BIC) and electric mirror (EM). However, the limitations of high-index materials are adverse to the practical applications, thus low-contrast metasurfaces with comparable performance are highly desired. Here we present a low-contrast dielectric metasurface comprising radially anisotropic cylinders, which are $SiO_2$ cylinders doped with a small amount of $WS_2$. The cylinder exhibits unidirectional forward superscattering originating from the overlapping of the electric and magnetic dipole resonances. When normal illumination by a near-infrared plane wave, the metasurface consisting of the superscattering constituents manifests a polarization-insensitive EM. Conversely, when subjected to an in-plane incoming wave, the metasurface generates a symmetry-protected BIC characterized by an ultrahigh $Q$ factor and nearly negligible out-of-plane energy radiation. Our work highlights the doping approach as an efficient strategy for designing low-contrast functional metasurfaces and sheds new light on the potential applications in photonic integrated circuits and on-chip optical communication.

## 1 Introduction

Bound state in the continuum (BIC) [1] is a fascinating and counterintuitive phenomenon in optics and photonics. Generally, a straightforward method for determining a bound state in the photonic crystal is to observe whether the location of its eigenfrequency lies outside the continuous spectrum, a region devoid of radiating pathways [2, 3]. However, in a departure from conventional understanding, the BIC as a localized bound state occurs within the radiation continuum, that is, the extended states exist across a continuous range of radiative frequencies [4, 5]. Thus the BIC is perfectly confined without any radiative loss yet possessing an ultrahigh quality factor $Q$. The formation mechanisms of BIC can be categorized into two classical types: symmetry-protected BIC at Γ point of the momentum space and accidental BIC off Γ point that results from destructive interference of two different resonances [6-9]. It should be noted that the BIC can also be present in acoustic [10, 11] and water waves [12, 13]. To date, BIC has been utilized extensively in fundamental research and technical applications [14], which encompass directive giant upconversion [15], distant interactions [16], harmonic wave [17], optical force

[18], surface-enhanced Raman scattering [19], gas detection [20], perfect absorption [21], nanolaser [22], beam shift [23], optofluidic transport [24], optical vortex beam [25], etc.

As we all know, a total reflection mirror is not only ubiquitous but also indispensable for manipulating light in optical and photonic systems. Traditional reflection mirrors are predominantly made of noble metals such as gold and silver, etc. However, these metal mirrors tend to experience unignorable energy loss or heat generation in the optical and near-infrared (NIR) regions due to plasmonic resonance and inherent material absorption [26-28]. Recently, an efficacious alternative has occurred in the form of all-dielectric metamaterial reflection mirrors with negligible absorption in these bands [29-32]. Intriguingly, one of the important branches is the perfect electric mirror (EM) functioning as a perfect electrical conductor (PEC), which exhibits total reflection along with a π-phase change of reflected electric field relative to the incident electric field at the reflection interface [32-35]. Consequently, the electric field at this interface demonstrates minimal intensity due to the destructive interference. The underlying physical mechanism of the EM originates from the Mie electric dipole (ED) resonance or the interferences of electric and magnetic multipoles [35-37]. Practically, a generalized EM is characterized by maintaining high reflection while the phase change ($\Delta\phi_E$) exceeds $\pm\pi/2$ along with the still weakening intensity at the interface [34, 38]. Therefore, by adjusting the geometry of each scatterer or the conditions of incidence, the realization of an EM with an adjustable $\Delta\phi_E$ becomes feasible due to the variable contributions of Mie multipoles [33, 37]. Currently, beyond supplanting the conventional metal reflection mirror in integrated optical and photonic systems, tunable EM also holds the potential for use as an optical switch [39].

To date, numerous phenomena involving BIC or EM have been explored in dielectric metasurfaces [19-21, 35, 36, 40], which are two-dimensional (2D) planar metamaterials. Utilizing the Mie resonances of dielectric particles, the metasurfaces can effectively manipulate the phase, amplitude, and polarization of the electromagnetic wave and have been applied in vector beam [41], generalized Brewster effect [42], data processing [43], etc. A remarkable feature of the metasurfaces associated with BIC [6, 15, 44-46] or EM [32, 33, 36, 47] responses is the high contrast of refractive indexes, typically, the high index scatterers but low index substrates and circumstances. Nonetheless, high-index materials have to contend with several constraints, including challenges in generating diverse patterns, performance degradation post-fabrication due to structural imperfections, and inhibiting efficient coupling between fibers and photonic circuits owing to polarization sensitivity [4, 48-50]. Therefore, it is meaningful to research the low-index dielectric functional metasurfaces. Although BIC has been implemented in low-index polymer, the overall structure with a high-index substrate retains the high contrast [51]. Inspiring, the accidental BIC has been observed in low-contrast dielectric grating composed of two materials with indexes very close [52]. However, to the best of our knowledge, symmetry-protected BIC and polarization-insensitive EM in low-contrast metasurface are rarely reported.

In this paper, we propose a NIR low-contrast metasurface comprised of radially anisotropic cylinders made of $SiO_2$ slightly doped with $WS_2$. Unlike the referenced work [52], the term 'low-contrast' in our study refers to the small refractive index difference between the cylinders to their surrounding medium air. Interestingly, the cylinder exhibits unidirectional forward superscattering resulting from the overlapped Mie mode resonances. As a result, the metasurface performs as a polarization-independent generalized total reflection EM when a normally incident plane wave, then, it generates symmetry-protected BIC when considering an in-plane incoming wave. Furthermore, transition metal dichalcogenides (TMDCs) have been applied to various BIC studies such as strong coupling [53], light emission [54], and electrical tuning [55]. Our work provides a strategy for manipulating the superscattering of low-index particles by doping anisotropic 2D materials and sheds new light on the potential applications of low-contrast metasurfaces in photonic circuits and on-chip optical communications.

## 2 Results and discussions
### 2.1 Superscattering for anisotropic cylinder

We begin by discussing the impact of radially anisotropic on the scattering properties of a low-index cylinder. An infinitely long isotropic $SiO_2$ cylinder embedded in the air is shown in Fig. 1(a), where the refractive index $n_l$ is 1.45 [56]. We assume an incident plane electromagnetic wave with wavevector **k** perpendicular to the center axis and magnetic field **H** parallel to the axis. Thus, according to Mie scattering theory [57], the $m$th-order scattering coefficient is written as

$$a_m = \frac{n_l J'_m(x_r) J_m(n_l x_r) - J_m(x_r) J'_m(n_l x_r)}{n_l J_m(n_l x_r) H_m^{(1)'}(x_r) - J'_m(n_l x_r) H_m^{(1)}(x_r)} . \quad (1)$$

Herein, $x_r$ is the size parameter calculated by multiplying the wavenumber $k$ with the radius $r$, $J_m(x)$ and $H_m^{(1)}(x)$ are the first kind of Bessel and Hankel functions respectively, and $f'(x)$ denotes the derivative of function $f(x)$ to $x$. Note that $m$ is an integer. Normalized by the single-channel scattering limit $2\lambda/\pi$ [58-60], the scattering efficiency reads as

$$N_{sca} = |a_0|^2 + 2\sum_{m=1}^{\infty} |a_m|^2 . \quad (2)$$

The total scattering efficiency of the cylinder as derived from Mie scattering theory (ST) is depicted in Fig. 1(b) by the blue curve. One can see that the homogeneous cylinder does not excite any appreciable scattering peaks. To verify the validity of the ST analysis results, we numerically calculate the total scattering efficiency via commercial software COMSOL based on the finite element method (FEM) [61], which is a well-suited approach for full-wave simulating electromagnetic responses of complex geometry shapes. The perfectly matched layer (PML) boundary is applied to substitute the outermost air for absorbing the scattered wave. The red circles (sim) in Fig. 1(b) show the total scattering efficiency based on the numerical simulation. The results represented by the circles coincide precisely with the curve, therefore, it confirms the validity of ST analysis and the response feature of no scattering peak.

Subsequently, we adopt the doping approach to improve the scattering efficiency of the cylinder. The structure schematic of the infinitely long radially anisotropic multiple layers (ML) cylinder is illustrated in Fig. 1(c). Here, the $SiO_2$ cylinder is doped with multilayer $WS_2$. To simplify, the $SiO_2$ and $WS_2$ layers are alternately arranged along the radial direction periodically. The total number of alternations is 15. The filling factor $f_1$ of the $WS_2$ layers is 0.077. This factor is determined by the ratio of the overall thickness of $WS_2$ layers to the diameter of the ML cylinder. The thickness is about 3.042 nm per $WS_2$ layer and 36.462 nm for $SiO_2$. With respect to a NIR plane wave with a wavelength $\lambda$ of 1550 nm in the air, $WS_2$ demonstrates anisotropic refractive indexes with a radial refractive index $n_r$ of 2.447 and an azimuthal refractive index $n_t$ of 3.756 according to the experiment measurement [62]. Note that the two indexes change very slowly in the NIR within $\lambda$ greater than 1 μm. Thus they are regarded as two constants with negligible absorption for our calculations [See Fig. S1 in Supplement 1].

To make the radially anisotropic work, incident electric field **E** is along the radial direction [63], which is the same as that in Fig. 1(a). Due to satisfying the long wave limit approximation, the effective medium theory (EMT) can be applied here. Based on the EMT [63-65], it is reasonable to approximate the radially anisotropic ML cylinder as a homogeneous one. The radial ($n_{rh}$) and azimuthal ($n_{th}$) refractive indexes are correspondingly expressed as

$$n_{th} = \sqrt{(1-f_1)n_l^2 + f_1 n_t^2} \ , \tag{3}$$

$$n_{rh} = n_l n_r \Big/ \sqrt{f_1 n_l^2 + (1-f_1)n_r^2} \ . \tag{4}$$

The anisotropy parameter is defined as $\eta = n_{th}/n_{rh}$, which is 1.169 for the equivalent cylinder. Then, according to the Mie scattering theory, the $m$th-order scattering coefficient of the equivalent cylinder can be written as [64, 66]

$$b_m = b_{-m} = \frac{n_{th} J_{\tilde{m}}(n_{th} x_r) J'_m(x_r) - J_m(x_r) J'_{\tilde{m}}(n_{th} x_r)}{n_{th} J_{\tilde{m}}(n_{th} x_r) H_m^{(1)'}(x_r) - H_m^{(1)}(x_r) J'_{\tilde{m}}(n_{th} x_r)} \ , \tag{5}$$

where $\tilde{m}$ is the radial anisotropy-revised function order $\tilde{m} = m\eta$. Therefore, the scattering efficiency normalized by the single-channel scattering limit is expressed in the same as Eq. 2, except that $a_0$ is substituted by $b_0$ and $a_m$ is interchanged with $b_m$, respectively.

Utilizing the aforementioned formulas, the total scattering efficiency of the isolated ML cylinder is demonstrated as the blue curve (ST) in Fig. 1(d). Compared to the ST curve depicted in Fig. 1(b), one can observe four prominent scattering peaks emerging in the vicinity of $x_r$=1.501, 2.402, 3.104, and 3.764, respectively. The scattering efficiencies at points $x_r$=1.501, 2.402, and 3.104 have enhanced from 0.602, 2.660, and 4.922 to 1.341, 3.873, and 5.169, respectively. Conversely, it has dropped from 7.030 to 5.259 at point $x_r$=3.764. Then, a detailed investigation of the scattering efficiency contributions of multipolar modes is available in section S2 of Supplement 1. The results elucidate the resonances of magnetic dipole (MD), ED, and electric quadrupole (EQ) modes lead to the discernible enhancement in the scattering efficiencies at the first three points of the previously mentioned $x_r$ values, respectively. Contrarily, the decline at the last point is attributed to the peak location separation of overlapping resonant MD, ED, and EQ modes. To validate the result of ST, the FEM is employed to explore the scattering responses of the ML cylinder. The red circles (sim) show the scattering efficiency of an actual radially anisotropic ML cylinder and the green curve (eff-sim) implies the scattering efficiency of the equivalent homogeneous cylinder using the effective material parameters. The numerical simulation results coincide nearly with the ST one, indicating slight doping $WS_2$ enables significant improvement in the scattering responses of the $SiO_2$ cylinder.

In the following, we focus on the scattering properties at $x_r$=2.402. Figure 2(a) provides the scattering efficiency contributions of multipoles of the ML cylinder. An efficient superscattering is achieved if the scattering efficiency magnitude is beyond the single-channel scattering limit [58, 64]. As the requirement of the rotational symmetry of cylinder coordination, the MD mode has only one scattering channel, while the remaining electric multipoles possess two scattering channels [36, 57, 64]. Therefore, the maximum value of the normalized scattering efficiency of MD mode equals one, while other electric modes peak at two. For convenience, point A denotes the $x_r$ of 2.402. Resonant MD and ED modes emerge simultaneously at point A due to the values of scattering efficiency being 1 and 2, respectively. Thus, a notable superscattering response of ED mode is captured owing to the magnitude beyond 1. Nevertheless, the scattering contribution of EQ mode cannot be negligible, whereas the much higher-order electric multipoles can be omitted safely. Essentially, the superscattering at point A originates from the interplay between the EQ mode and the concurrently resonant ED and MD modes. It is worth emphasizing that the location parameter of overlapping resonances of the ML cylinder is much smaller than the pure cylinder. Reduction of size is convenient for device minimization and integration.

To assess the far-field properties of the superscattering, the angular scattering distribution of the ML cylinder at point A is investigated. Utilizing the accessible multipolar scattering coefficients, the angular scattering intensity is written as [57, 64]

$$I_{SA}(\theta) = \frac{2}{\pi k}\left|b_0 + 2\sum_{m=1}^{\infty} b_m \cos(m\theta)\right|^2, \qquad (6)$$

where $\theta$ is the scattering angle and the forward and backward directions correspond to the 0 and 180 degrees, respectively. The angular scattering distributions of distinct multipoles and their total term obtained by the interferences of the multipoles are depicted in Fig. 2(b). Accordingly, the angular distribution of the total term (sca, red curve) indicates the unidirectional scattering at point A, namely the forward scattering is enhanced significantly but backward scattering is suppressed approaches to zero. To quantify the unidirectional degree, an asymmetry parameter is defined as [31, 57]

$$g = \frac{\oint_s \cos\theta W_{sca}(\theta)ds}{\oint_s W_{sca}(\theta)ds}, \qquad (7)$$

where $W_{sca}(\theta)$ is the scattering power, and $s$ is a surface enclosing the particle. In particular, $g=0$ is the light scattering isotropically or laterally, $g=1$ is the perfect forward scattering, and $g=-1$ is perfect backward scattering. As a result, the $g$ of point A is 0.783. To further test the directional, a parameter written as $\beta$ is defined by the forward-to-backward scattering intensity ratio. Here, the $\beta$ is 37.109. Thereby, the ML cylinder manifests good unidirectional forward superscattering.

Subsequently, Fig. 2(c) illustrates the scattering electric field $|\mathbf{E}_s|$ distribution of the ML cylinder at point A using effective material parameters. Accompanied by the incident wavevector $\mathbf{k}$, the energy primarily concentrates inside the cylinder. However, the scattered energy outside of the cylinder mainly presents in the forward direction, while the backward energy approximates to zero. On the other hand, Fig. 2(d) portrays the $|\mathbf{E}_s|$ distribution of the actual radially anisotropic ML cylinder. Comparatively, the distributions in Figs. 2(c) and 2(d) are almost identical. Therefore, the numerical simulations are consistent with the result of the angular scattering profiles, verifying the occurrence of the unidirectional forward superscattering of the ML cylinder at point A.

**2.2 Polarization-insensitive electric mirror**

In what follows, we propose a dielectric metasurface possessing low contrast of the refractive indexes. Figure 3(a) shows the configuration of the metasurface composed of identical radially anisotropic ML cylinders with radii of 592.550 nm corresponding to the value at point A. The cylinders are infinitely arranged along the $x$-direction with a period of $a$. The entire metasurface is orthogonal to the $y$-direction and immersed in air. Employ the previously described transverse electric (TE) plane wave characterized by a wavelength of 1550 nm and an $x$-polarised $\mathbf{E}$, to normally impinge on the metasurface. As a consequence, each ML cylinder excites the unidirectional forward superscattering.

As a function of $a$, Fig. 3(b) shows the simulated results of the reflectivity (blue curve, eff) and $\Delta\phi_E$ (red curve, eff) of the metasurface composed of the ML cylinders adopting effective material parameters. Here, $\Delta\phi_E$ is normalized by $\pi$. From the blue curve, a prominent phenomenon of total reflection emerges at point B. Point B stands for $a$=1.506 μm. Referring to the grating diffraction theory [67], the total reflection is the zeroth order wave because the period is less than the normal incidence wavelength. Furthermore, the $\Delta\phi_E$ at point B is 0.874$\pi$, which far exceeds $\pi/2$ and closes $\pi$. Consequently, a generalized EM emerges at point B. Essentially, the EM response is associated with the interferences between the dominant ED resonance and the other multipoles of the superscattering cylinder in accordance with the multipolar scattering analysis in Fig. 2(a). In addition, the period dependence of the blue and red curves in Fig. 3(b) indicates the lattice size also plays a key role in the EM response. On the other hand, the reflectivity and $\Delta\phi_E$ of the metasurface consisting of the actual ML cylinders are exhibited as blue and red circles, respectively. In this circumstance, the reflectivity is 0.988 and $\Delta\phi_E$ is $0.836\pi$. The two values are in good agreement with those of the effective material parameters model. Therefore, the metasurface made of the actual radially anisotropic ML cylinders behaves as an identical EM response.

Figure 3(c) shows the $|\mathbf{E}_x|$ distribution at point B of one unit cell of the metasurface with effective material parameters. The significant standing wave interference fringes stemming from the incident and reflected electric fields occur in the reflection space, while the fringes are absent in the transmissive space. It validates the metasurface excites total reflection, which meets the first key feature of the EM. Then, focusing on the localized field, the fringe intensity of destructive interference on the top interface of the metasurface is near the minimum, which satisfies the second key feature of the EM. Subsequently, the pronounced electric field energy confinement between the neighboring cylinders manifests the occurrence of the waveguide-array (WGA) mode [68, 69]. For comparison, Fig. 3(d) depicts the $|\mathbf{E}_x|$ distribution at point B of the metasurface composed of actual ML cylinders. Ignoring the difference inside the cylinder, the external $|\mathbf{E}_x|$ distribution is in concordance with that in Fig. 3(c). Therefore, the response consistency between the effective parameter and actual geometry models confirms the EM response originating from the interactions of the unidirectional forward superscatterings and the WGA mode.

Next, we will evaluate the impact of incident polarization on the EM. The validity of the effective parameter model is confirmed by the previous discourse, thus, our forthcoming analyses will concentrate on this model for the sake of brevity. Figure 4(a) shows the reflectivity (blue curve) and $\Delta\phi_E$ (red curve) of the same configuration metasurface as in Fig. 3(a) illuminated by the same incident wave except for the only difference of $z$-polarized $\mathbf{E}$, namely a transverse magnetic (TM) wave. At point B, the reflectivity is unity apparently and $\Delta\phi_E$ is $-0.653\pi$ satisfying destructive interference. Figure 4(b) demonstrates the $|\mathbf{E}_z|$ distribution of one unit cell. According to the interference fringes, the occurrence of destructive interference at the top interface of the cylinder is confirmed. Hence, the metasurface still exhibits the EM response at point B. The salient localized electric field energy between adjacent cylinders indicates the presence of the WGA mode. Nonetheless, the excited specific scattering modes are changed due to the distinct electric field patterns near and inside the cylinder compared to Fig. 3(c).

To clarify the excited scattering modes, we investigate the scattering properties of an isolated ML cylinder. Correspondingly, the $m^{\text{th}}$-order scattering coefficient of the equivalent cylinder can be written as [66]

$$c_m = c_{-m} = \frac{n_{th}J'_m(n_{th}x_r)J_m(x_r) - J'_m(x_r)J_m(n_{th}x_r)}{n_{th}J'_m(n_{th}x_r)H_m^{(1)}(x_r) - H_m^{(1)'}(x_r)J_m(n_{th}x_r)} \quad . \tag{8}$$

Similarly, the expression formation of the normalized scattering efficiency is identical to Eq. 2, while $a_0$ is replaced by $c_0$ and $a_m$ is interchanged with $c_m$, respectively. As a result, Fig. 4(c) shows the scattering efficiency contributions of the multipole modes based on the formulas above. It demonstrates that the cylinder at point A primarily supports MD resonance along with other electric and magnetic multipolar modes. Notably, both MD and MQ intensities surpass unity, suggesting that the cylinder continues to exhibit superscattering behavior. Compared with Fig. 2(b), the variation in $\Delta\phi_E$ for the two EMs arises from the transition of excited dominant resonance from ED to MD modes and the contribution variation of other multipolar modes. Additionally, the angular scattering spectra of the individual cylinder are illustrated in Fig. 4(d). The interferences between the ED, MD, MQ, and MO modes lead to the total scattering denoted by the red curve. Although the backward is slightly enhanced compared with Fig. 2(b), the $g$ is 0.742 and the $\beta$ is still as high as about 11.363. Figure 4(e) shows the $|\mathbf{E}_s|$ distribution of the cylinder at point A under the incident TM wave with wavevector $\mathbf{k}$. The majority of energy is confined in the forward direction. Consequently, the cylinder exhibits remarkable unidirectional forward scattering. By now, the precise understanding of the origination of the EM can be attributed to the interaction of the unidirectional forward superscatterings of the cylinders and the WGA mode of the lattice. Consequently, keeping the superscattering and WGA mode, the metasurface behaves as the polarization-insensitive EM.

**2.3 Band state in the continuum**

In this section, we will investigate the BIC based on the same metasurface at point B in Fig. 3(a). In a special scenario of only considering the in-plane incident wavevector $\mathbf{k}=k_x\mathbf{e}_x$, $\mathbf{e}_x$ represents the unit vector along the $x$ direction, the metasurface is treated as a one-dimensional photonic crystal open in the $y$ direction. By numerical calculation, Fig. 5(a) describes the normalized TE band diagram in the first Brillouin zone, i.e. lower band $TE_1$ of red and upper band $TE_2$ of blue curves respectively. Noteworthy that the band is selected only when its corresponding energy at Γ point is primarily localized in the metasurface, otherwise, the pseudo band is omitted. Since the entire metasurface is embedded in the air, the angular frequency $\omega$ of the modes lying along the light line can be derived as $\omega(k)=ck_x$ [2]. Accordingly, all modes along the $TE_1$ and $TE_2$ bands are above the light line, which suggests the two bands are located in the continuum.

Then, dependent on the band diagram analysis, a specific mode can be associated with a complex eigenfrequency, $f=f_0-i\gamma$. Here, the real part $f_0$ is the resonant frequency and $\gamma$ denotes the energy leakage rate. We emphasize that the total leakage rate $2\gamma$ here can be replaced by total radiative loss $\gamma_r$ due to the non-radiative loss being forbidden when the material is intrinsic loss-free [44]. Thus, the corresponding radiative quality factor $Q_r$ can be expressed as [44, 45], $Q_r= f_0/\gamma_r$. Accordingly, the $Q_r$ of $TE_1$ and $TE_2$ bands are respectively computed as the red and blue curves in Fig. 5(b). The maximum $Q_r$ of the two bands occurs synchronously at Γ point, where the $Q_r$ of the $TE_1$ band is near $10^9$ far greater than the $TE_2$ band. To achieve the bound state, a relatively high $Q_r$ is one of the crucial features, thus we will focus on the $TE_1$ band in the following.

Figure 5(c) indicates the normalized total radiative loss of the $TE_1$ band. The radiative loss reaches a minimum at Γ point, where the value approaches zero. It reveals that the metasurface at Γ point rarely radiates energy outwards. Moreover, the metasurface is symmetric about up-down (i.e. $+y$ and $-y$ directions in Fig. 3(a)) and left-right. Specifically, the structure is infinitely periodically arranged in the left-right direction, yet open in the up-down direction. Thus the in-plane energy is confined and the energy radiative channels only appear in upward and downward directions. Thanks to the up-down symmetry, the radiative loss towards the top $\gamma_t$ is always equal to that towards the bottom $\gamma_b$ theoretically [44, 45]. So, $\gamma_r=\gamma_t+\gamma_b$. The normalized $\gamma_t$ and $\gamma_b$ are shown in Figs. 5(d) and 5(e), respectively. The behaviors of upward and downward radiative loss are synchronous as well as at Γ point exhibits almost non-radiative, whereas the other points are significant.

To validate the preceding radiative loss properties, the far-filed energy density at the top ($W_t$) and the bottom ($W_b$) are presented in Figs. 5(f) and 5(g), respectively. These outcomes are in concordance with the results of radiative loss, namely the $W_t$ and $W_b$ display the same trend, likewise, the upward and downward far-field energies at Γ point are far lower than the other points. Thus, the metasurface at Γ point hardly radiates energy. Moreover, the **E** norm distribution of one unit cell at Γ point is depicted in Fig. 5(h). One can see the **E** energy is predominantly confined in the in-plane of the metasurface, including the inside of cylinders and the gaps of adjacent cylinders. Then, there is no **E** energy distribution in the far-field regions at the top and the bottom. It is noteworthy that the $x_r$ of the cylinder at Γ point of the $TE_1$ band is 2.023, which implies the isolated cylinder excites the dominant ED mode as well as the MD resonance according to Fig. 2(a). The cylinder maintains the superscattering due to the intensity of ED mode beyond the single-channel limit (more details see S3 of Supplement 1). Factually, due to the strong excitation of multipolar modes, we can observe a small amount of **E** energy leaking into the upward and downward near-field air regions. Additionally, the **H** norm distribution of one unit cell at Γ point is demonstrated in Fig. 5(i). Similarly, the majority of **H** energy is concentrated inside the cylinder, meanwhile, a minor leak into the nearby air regions. The leaked electromagnetic energy hinders attaining an infinite $Q_r$ factor. Notwithstanding, the in-plane confinement of electromagnetic energy and the non-radiation in the far-field indicate the mode at Γ point of the $TE_1$ band is a BIC.

To further understand the physical picture of BIC, we delve into discussing the responses at Γ point as the period varies. Figure 6(a) shows the $Q_r$ response and the $TE_1$-Γ marks the Γ point of the $TE_1$ band in Fig. 5(a). The $Q_r$ increases slowly with the period increasing, however, all $Q_r$ maintain the same order of magnitude exceeding $10^8$. On the contrary, as depicted in Fig. 6(b), the normalized radiative loss decreases slowly with an increase in the period, while all orders of magnitudes are about $10^{-9}$. Therefore, despite the period variation, the BIC with ultrahigh $Q_r$ and extremely low radiation loss persist in the metasurface and exhibits good robustness. Therefore, we can sufficiently prove the bound state at the $TE_1$-Γ is indeed a symmetry-protected BIC since the period change fails to destroy the structure symmetry [44-46]. Figure 6(c) depicts that the eigenfrequency of the BIC in the NIR region experiences a monotonic redshift with the period increasing.

**Conclusion**

In summary, we have constructed a low-contrast metasurface consisting of radially anisotropic ML cylinders. These structures are formed by doping a small amount of $WS_2$ into the low refractive index $SiO_2$ cylinders. By comparison, we verify the EMT is suitable enough for our research due to the thickness of each layer of the ML cylinder being far smaller than the effective wavelength. Consequently, the doped ML cylinder significantly improves the scattering responses compared to the pure $SiO_2$ cylinder. As a result, we archive a unidirectional forward superscattering with a much smaller size owing to the overlapped ED and MD resonances. When illuminated by a normal incident wave with a wavelength of 1550 nm, the metasurface performs as a polarization-insensitive EM, which stems from the interactions of the unidirectional forward superscatterings of the cylinders and the WGA mode of the lattice. Moreover, when taking in-plane excitation into account, the metasurface exhibits a symmetry-protected BIC characterized by ultrahigh $Q$ and extremely low radiative loss leading to energy non-radiation in far-field and strong energy confined in-plane.

Noteworthy to mention that the Kerker scatterings [31, 70], stemming from the interaction of ED and MD modes with identical magnitude, are also able to generate unidirectional either forward or backward scattering. However, they may not necessarily form superscattering. Our work provides a new feasible strategy to promote the scattering response of a low-index isotropic object via doping anisotropic layered TMDCs including $WS_2$, $WSe_2$, $MoS_2$, and $MoSe_2$ [62], etc. Then, the proposed low-contrast metasurface is promising to address the partial limitations of high-contrast metasurfaces [4]. Specifically, it enables performance as a polarization-insensitive total reflection EM to boost the coupling of the photonic integrated circuits to fibers [48] and offers a potential platform to achieve unidirectional guided resonances via breaking the up-down mirror symmetry leading to enhancement of the directional one-side coupling efficiency in photonic devices [44, 71, 72]. Next, through doping TMDCs and utilizing their excitonic polarization [73], we will manipulate the optical scattering of low-index particles, then design BIC in a low-contrast metasurface to investigate the nonlinear effects in photonic devices.

**Acknowledgments**

This work was supported by the National Natural Science Foundation of China (Nos. 12304425, 12074267, 12374355), Key project of National Key Research and Development Program of China (2022YFA1404500).

**Disclosures**

The authors declare no competing financial interest.

**Data availability.** Data may be obtained from the authors upon reasonable request.

**Supplemental document.** See Supplement 1 for supporting content.

**Figures**

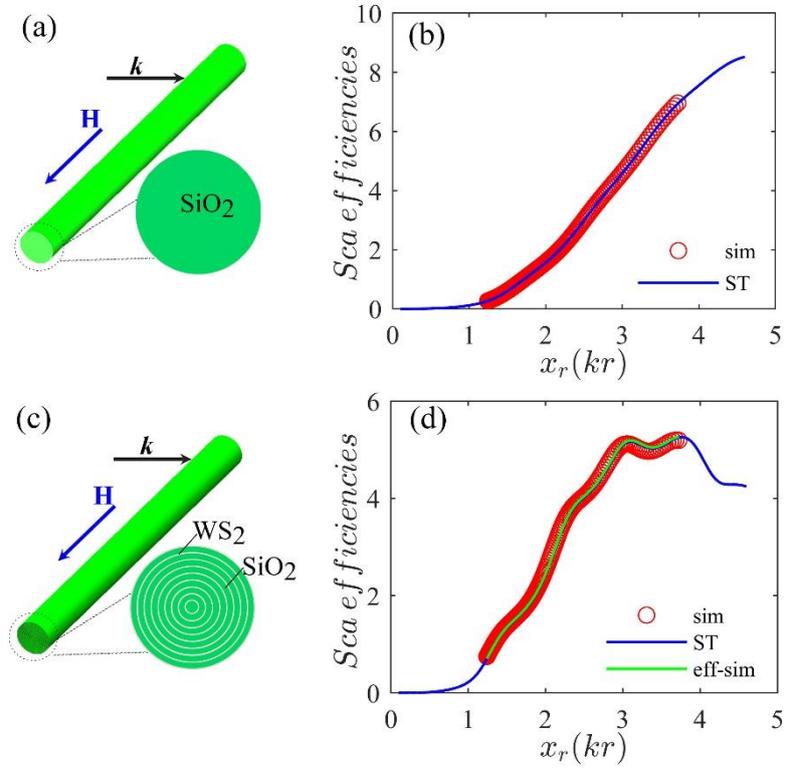

Figure 1. Scattering properties of an infinite-long cylinder. (a) Schematic of an individual $SiO_2$ cylinder under the normally incident plane wave with incident wavevector ***k*** and magnetic field **H** paralleling the center axis. (b) Total scattering efficiencies of the cylinder. Size parameter $x_r$ obtained by multiplying wavenumber $k$ and radius $r$. Red circles were obtained by the numerical simulation (sim), and the blue curve was calculated by the scattering theory (ST). (c) configuration of a single multiple-layer (ML) cylinder illuminated by the same imping wave. (d) Total scattering efficiencies of the ML cylinder. The green curve (eff-sim) was obtained by numerical simulation with the effective material parameters.

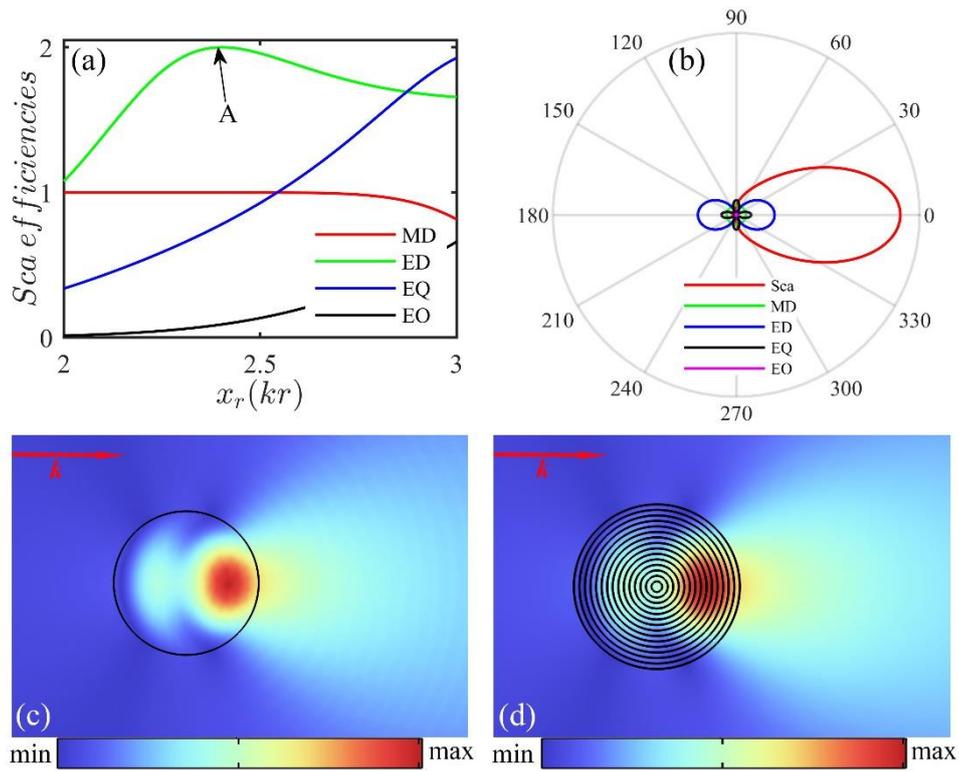

Figure 2. Superscattering analysis of the ML cylinder. (a) The multipolar scattering efficiencies of the ML cylinder via the scattering theory. Point A denotes the $x_r$=2.402. (b) Angular scattering of the ML cylinder at point A. (c) Far-field scattering electric field |**E**$_s$| at point A obtained by the numerical simulation with effective material parameters. ***k*** is the incident wavevector. (d) |**E**$_s$| of the actual ML cylinder at point A.

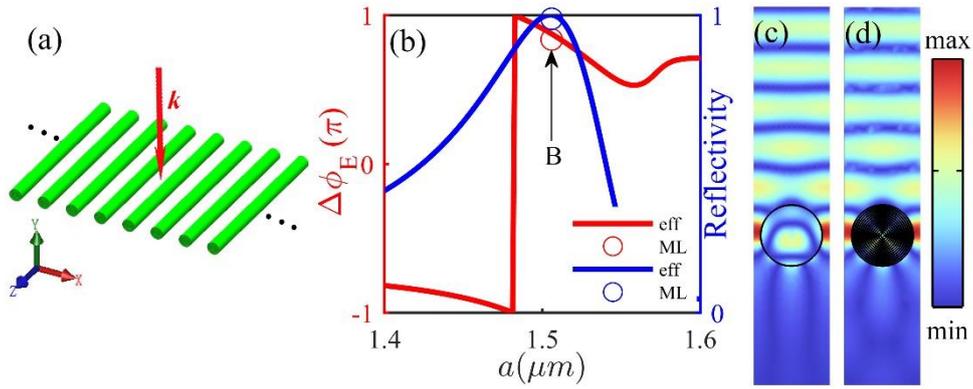

Figure 3. EM responses of low-contrast metasurface under a normally incident TE plane wave, i.e. with **k** and *z*-polarized **H**. (a) Geometry of the low-contrast metasurface composed of the radially anisotropic ML cylinders. The lattice period is *a*. (b) Simulation results of reflectivity and $\Delta\phi_E$. The curves were obtained with effective material parameters, and the circles were the case of the actual ML cylinders. Point B designates the *a*=1.50595 μm. (c) Electric field |**E**$_x$| distribution of one unit cell of the metasurface with effective material parameters. (d) |**E**$_x$| distribution in the case of the actual ML cylinders.

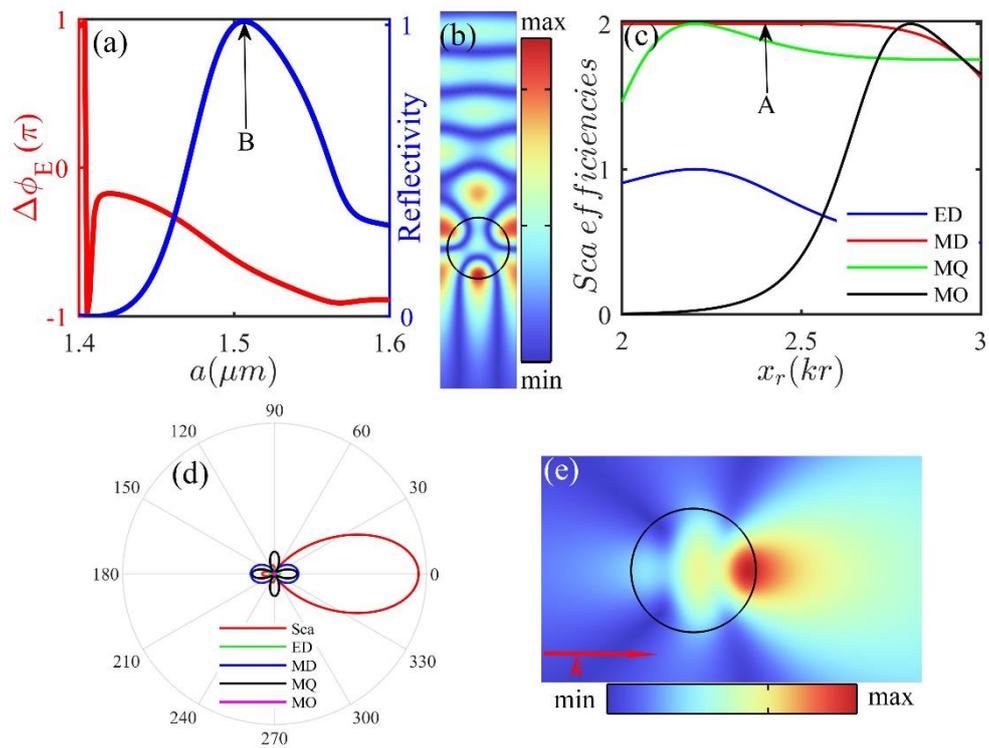

Figure 4. EM responses of the low-contrast metasurface under a normally incident TM plane wave, i.e. $z$-polarized **E**. (a) Simulation results of reflectivity (blue) and $\Delta\phi_E$ (red). (b) Electric field $|E_z|$ distribution at point B. (c) Scattering efficiencies of the ML cylinder calculated by the scattering theory. (d) Angular scattering pattern of point A. (e) $|\mathbf{E}_s|$ distribution of point A.

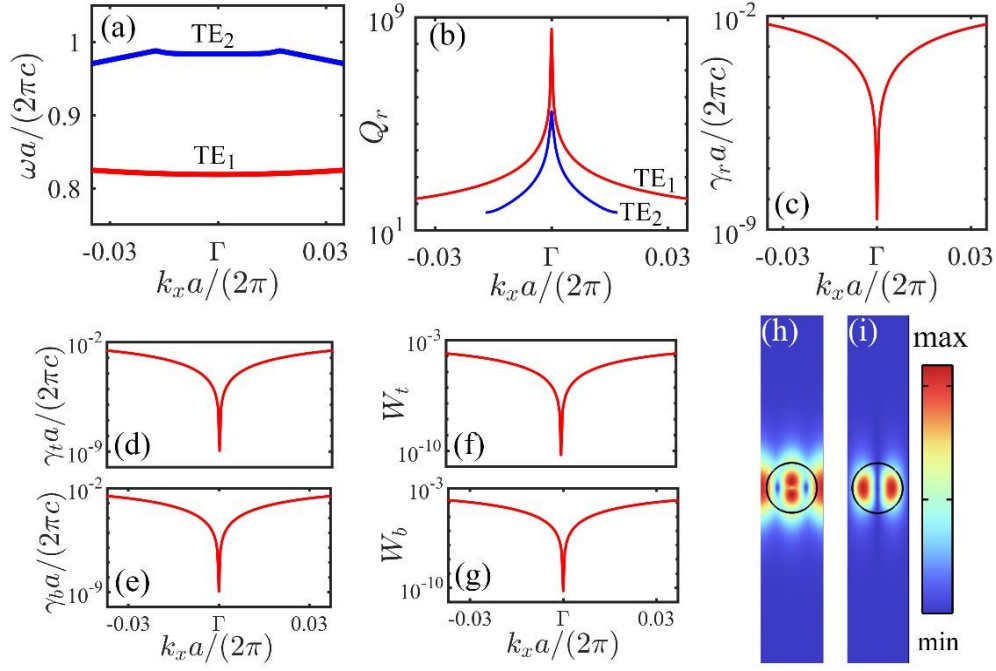

Figure 5. BIC properties of the low-contrast metasurface illuminated by TE wave with in-plane wavevector. (a) Calculated TE band diagram of the metasurface at point B. (b) Radiative quality factor of $Q_r$ for the $TE_1$ (red curve) and $TE_2$ (blue curve) bands. We focus on the $TE_1$ band. (c) Total radiative loss. (d) Radiative loss toward the top. (d) Radiative loss toward the bottom. (e) Radiation energy toward the top ($W_t$), unit is J/m². (f) Radiation energy toward the bottom ($W_b$), unit is J/m². (h) Electric field norm distribution of the Γ point. (i) Magnetic field norm distribution of the Γ point.

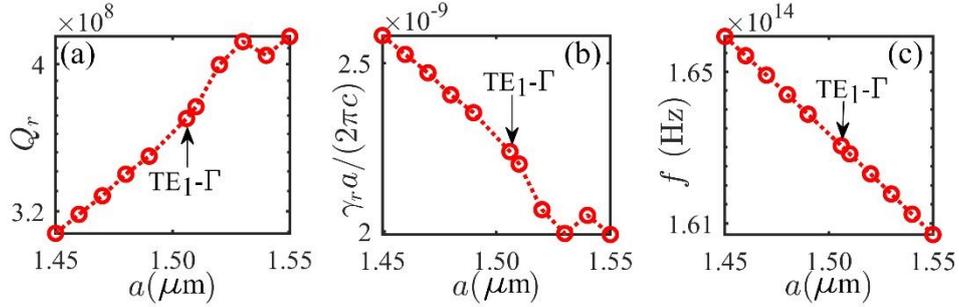

Figure 6. BIC responses as a function of period $a$, (a) the $Q_r$, (b) the normalized radiative loss, and (c) the eigenfrequency $f$, respectively. $TE_1$-Γ denotes the Γ point of the $TE_1$ band in Fig. 5(a).